\documentclass[aps,prc,twocolumn,groupedaddress,superscriptaddress]
{revtex4}
\usepackage[utf8]{inputenc}
\usepackage{graphicx} 
\usepackage{dcolumn} 
\usepackage{longtable} 
\usepackage{amssymb} 
\usepackage{amsmath} 
\usepackage{amsfonts} 
\usepackage{slashed} 
\usepackage[dvipsnames]{xcolor}
\usepackage{xfrac}
\usepackage{subfigure}
\usepackage[hidelinks]{hyperref}
\newcommand{\bs}[1]{\boldsymbol{#1}} 
\newcommand{\nopieft}{\mbox{$\slashed{\pi}$EFT~}} 
 
\newcommand{\Lag}{{\cal L}} 
\newcommand{\be}{\begin{equation}} 
\newcommand{\ee}{\end{equation}}

\newcommand{\pvec}{{\bs{p}}}

\newcommand{\reff}{r_{\text{eff}}}
\graphicspath{{figures/}}
\begin{document}
\title{Removing the Wigner bound in non-perturbative effective field theory}

\author{Saar Beck}
\affiliation{The Racah Institute of Physics, The Hebrew University, 
Jerusalem 9190401, Israel}
\author{Betzalel Bazak}
\affiliation{The Racah Institute of Physics, The Hebrew University, 
Jerusalem 9190401, Israel}
\author{Nir Barnea}
\affiliation{The Racah Institute of Physics, The Hebrew University, 
Jerusalem 9190401, Israel}

\date{\today}

% ########
% ABSTRACT
% ########
\begin{abstract}
    The Wigner bound, setting an upper limit on the scattering effective range,
    is examined at different orders of contact effective field theory. Using
    cutoff regulator we show that the bound loosens when higher orders of the
    theory are considered.
    For a sharp and a Gaussian regulators, 
    we conjecture an analytic 
    formula for the  dependence of the Wigner bound on the theory's order. 
    It follows that the bound vanishes in the limit of infinite order.

    Using a concrete numerical example we demonstrate that the above surmise 
    still holds after renormalization
    at finite cutoff.
    Studying the 3-body system with this example, 
    we have found that limiting the permissible range 
    of cutoffs by the Wigner bound, we avoid the
    Thomas collapse, and don't need to promote the 3-body force to leading order.

%    Furthermore,
%    we find that there exist multiple renormalization choices, only one of
%    which is physical. A method to pick the physical solution is suggested.
\end{abstract}

\maketitle

% =============================================================================
% INTRODUCTION 
% =============================================================================

\section{Introduction}
In the last two decades contact effective field theory (EFT) has been 
successfully applied for studying low energy systems where the
characteristic particle wave length is much larger than the interaction range.
Such systems can be found, for example, in ultra cold atomic gases, clusters of 
He atoms, and atomic nuclei, 
see e.g. \cite{Higa2008,Braaten2006,Hammer2017,Hammer2019}.
In such circumstances, it can be argued that the
details of the short-range part of the interaction become irrelevant 
and a contact interaction, i.e. a delta function and its derivatives, 
can be used to parameterize it \cite{Lepage1997,Bedaque2002,Beane2000}.
This parameterization can be understood to result
from integrating out the high energy
degrees of freedom of the underlying theory and then expanding
the resulting Lagrangian in terms of contact
interactions.

In a complete EFT, the Lagrangian should include all possible terms compatible
with the symmetries of the underlying theory.
To make such theory of any use, only finite number of terms should be retained 
when aiming to calculate some observable to a desired accuracy.
Therefore an appropriate ordering of the Lagrangian terms is a key ingredient
for a successful EFT. A natural order for contact EFT is arranging the
interaction terms according to their mass dimension
\cite{Kaplan2005,Hammer2019,Epelbaum2017}.
A non-relativistic particle field $\Psi$ counts as $3/2$ and a derivative
counts as $1$. This ordering is known as the ``na{\"iv}e'' power counting.
It works very well for the 2-body sector if the scattering length 
$a_s$ is of the order of the interaction range, 
but fails for very large scattering lengths (see e.g. \cite{PavonValderama2017} for a nice
derivation), 
and for the $3,4,\ldots$-body,  
interaction terms \cite{BHvK99,Bazak2018a}.

In the scheme proposed by Weinberg \cite{Weinberg1990}, {the contact
interactions are to be iterated in the Lippmann-Schwinger equation to yield
the $T$-matrix.}
Analyzing the ability of such a scheme to reproduce the low energy $s$-wave
effective range expansion, the scattering length and
effective range $\reff$, Phillips {\it et al.} \cite{Phillips1997,Phillips1998}
have argued, using cutoff regulation, that the resulting
effective range is necessarily negative in the limit of infinite cutoff,
$\Lambda\to\infty$.
Furthermore, analyzing specifically a contact EFT at next-to-leading-order (NLO)
with a 2-body potential consisting of a delta function plus its
derivative-squared, they have shown that in the $\Lambda \to \infty$ limit
$\reff \leq 0$ regardless of the Lagrangian parameterization, i.e. the value
of the low energy constants (LECs). 

The deep fundamental reason behind this observation can be traced back to
the causality principle. For a scattering process, causality implies that
the scattered wave cannot leave the target before the incident wave has
reached it. Applying this argument to 2-body scattering with an energy
independent potential of range $R$, Wigner has shown \cite{Wigner1955},
followed by others \cite{Hammer2010,Phillips1997}, that the effective range
has an upper bound,
\be\label{eq:wigner_bound}
  \reff\leq 2R\left(1-\frac{R}{a_s}+\frac{R^2}{3a_s^2}\right)\;.
\ee
The range $R$ of the interaction is proportional
to $1/\Lambda$. Hence, the Wigner bound becomes more restrictive with 
increasing $\Lambda$. 
In the limit of $\Lambda \to \infty$ ($R\to 0$), the effective range is bounded 
from above by zero \cite{Phillips1998}.

This bound has a profound effect on contact EFT. The theory cannot be
renormalized while working non-perturbatively \cite{Beane1998}. 
More precisely, the cutoff cannot be eliminated without turning $\reff$ to be
negative, if any order but the leading order is treated non-perturbatively
\cite{vanKolck1999}. 
Consequently, two options remain, either working non-perturbatively giving up 
renormalization group invariance, see e.g. \cite{Bansel2018,Lensky2016}, 
or working perturbatively for any EFT order beyond leading order, 
see e.g. \cite{Hammer2017}.
The latter option is much harder numerically for the many-body problem,
since it requires resolving the full leading order Green's function.

Here we would like to take a second look at non-perturbative contact EFT, and
check if it is forever doomed to be cutoff dependent or maybe renormalization
group invariance can be restored if we include enough orders in our calculation.
Put differently - what will happen if we first take infinite EFT orders and
only then push the cutoff to infinity? 
As a motivation consider the EFT as a sum of delta functions and their
derivatives \cite{Lepage1997}, 
generated from a Gaussian with a width of $\frac{1}{\Lambda}$. 
In the limit $\Lambda\to \infty$ the delta function is recovered.
For higher EFT orders the interaction 
range is associated with
terms of the form $x^n\exp(-x^2/2)$,
i.e. higher order terms broaden the interaction range.
For a specific order the resulting range is finite and by taking
$\Lambda\to\infty$ the range vanishes. But this may not be the case for infinite
EFT order where there are infinite number of terms which broadens the
interaction range. 

To study this point we consider here a contact EFT of scalar bosonic
fields. First we show explicitly using partial renormalization
that up to order N$^9$LO,
given a scattering length, the Wigner
bound loosens as more EFT orders are taken into account. Then we demonstrate, 
for a concrete example at N$^2$LO, that our finding holds when the LECs
are renormalized to reproduce the effective range expansion to order $p^4$.  
From this analysis we conjecture that for an arbitrary value of $\Lambda$, any 
finite effective range $\reff$ can be described by the theory if the EFT order
is large enough. 
In the process, we find that multiple renormalization choices arise as more
orders are taken into account. A method to pick the physical one is then
suggested.

% ################################################
\section{Effective range in increasing EFT orders}
\label{Effective range in various EFT orders}
% ################################################

In order to understand the evolution of the Wigner bound with increasing 
EFT orders, we study a low energy EFT for a scalar bosonic 
field $\Psi$ of mass $m$. 
This theory is closely related to the nuclear pionless effective field theory
(\nopieft), that as fundamental degrees of freedom includes only nucleons,
with no mesons. 
Starting from the EFT Lagrangian $\Lag$, and deducing the 2-body interaction,
the low energy scattering parameters are obtained by solving the $s$-wave
Lippmann-Schwinger equation. Expanding the resulting $T$-matrix in the usual
low-momentum form
\begin{align}\label{eq:t_expansion}
  \frac{1}{T} =
   -\frac{m}{4\pi}\left(-\frac{1}{a_s}+\frac{1}{2}\reff~ p^2
   +\sum_{n\geq 2} S_n p^{2n}-ip\right),
\end{align}
the scattering length $a_s$ is then identified as the leading momentum
independent term, the effective range $\reff$ as the energy $p^2$ coefficient,
and the shape parameters $S_n$ through the $p^{2n}$ terms.

Considering only $s$-wave interactions, and ignoring 3-body and higher
multi-particle interaction terms, the EFT Lagrangian can be written as
\be\label{eq:lagrangian}
 \Lag = \Lag_0+\Lag_1+\Lag_2+\ldots
\ee
where $\Lag_0$, the LO Lagrangian, includes the free Lagrangian and a
contact interaction term
\be\label{eq:lag_0}
  \Lag_0 = \Psi^{\dagger}\left(i\partial_t+\frac{1}{2m}\nabla^2\right)\Psi
          -\frac{1}{4}C_{00}\left(\Psi^{\dagger}\Psi\right)^2\;,
\ee
$\Lag_1$ includes the subleading NLO interaction term
\be\label{eq:lag_2}
 \Lag_1 =  -\frac{1}{8}{C_{20}}\left[\left(\Psi^{\dagger}\Psi\right)
                              \left(\Psi^{\dagger}\nabla^2\Psi\right)
          +h.c.\right]\;,
\ee
and $\Lag_2$ the N$^2$LO interaction term
\begin{align}\label{eq:lag_2}
  \Lag_2 = &
    -\frac{1}{8}{C_{40}}\left[\left(\Psi^{\dagger}\Psi\right)
                       \left(\Psi\nabla^4\Psi^{\dagger}\right)+h.c.\right]
    \cr &
  -\frac{1}{4}C_{22}\left(\Psi^{\dagger}\nabla^2\Psi\right)
                   \left(\Psi^{\dagger}\nabla^2\Psi\right)\;.
\end{align}
In a similar fashion, $\Lag_{n}$ includes all possible 4-field operators with
$n$ insertions of $\nabla^2$. The Lagrangian parameters $C_{pq}$, known as the
low energy coefficients (LECs), are fixed through the renormalization condition
to reporoduce the scattering parameters $a_s,\reff,S_n$, etc. For brevity, we
also write $C_{00}$ as $C_0$, $C_{20}$ as $C_2$, and so on.
 
In momentum space, the corresponding potential can be written as 
\begin{align}\label{eq:vnlo}
  V_{\text{N}^n\text{LO}}=&C_0+C_2\left(p^2+p'^2\right)
  \cr & \hspace{1em}+
  C_4\left(p^4+p'^4\right)+C_{22}p^2p'^2+\cdots\;,
\end{align}
where $\pvec$ is the incoming relative momentum and $\pvec'$ the outgoing
momentum. To avoid the UV divergences embedded in the contact terms we 
regularize the interaction $V \to F(\pvec')V F(\pvec)$ using
either a Gaussian regulator $F(p) = e^{-\frac{p^2}{\Lambda^2}}$
or a sharp cutoff $F(p) = \Theta(\Lambda-p)$, where $\Theta(\Lambda-p)$ is the
Heaviside step function.
    
The $T$-matrix is obtained by iterating the potential through the
Lippmann-Schwinger equation
\be
    T=V+VGT\;.
\ee
Following the footsteps of Phillips, Beane and Cohen \cite{Phillips1998},
we note that for a separable potential such as the EFT interaction
\eqref{eq:vnlo}, which can be written as
\be\label{eq:lambda}
  V(p,p') = F(p)\sum_{i,j=0}^{n} p^{2i}\lambda_{ij} p'^{2j} F(p')\;,
\ee
the $T$-matrix assumes the form
\be\label{eq:t_matrix}
  T(E) = F(p)\sum_{i,j=0}^{n} p^{2i} \tau_{ij}(E) p'^{2j} F(p')\;,
\ee
and the Lippmann-Schwinger equation is reduced into a simple matrix equation
\be\label{tau_eq}
  \tau(E) = \lambda+\lambda {\cal I}(E) \tau(E)
\ee
where the matrix elements of ${\cal I}$ are given by
\be\label{def_I}
   {\cal I}_{ij} \equiv 
        \int\frac{d^{3}q}{(2\pi)^{3}}
             \frac{F^2(q) q^{2(i+j)}}{E+i\varepsilon-\frac{q^2}{2\mu}}\;
\ee\
and $\mu=m/2$ is the reduced mass.
Equation \eqref{tau_eq} can be formally solved to yield
\begin{align}\label{tau_solution}
  \tau(E) = \frac{1}{1-\lambda{\cal I}(E)}\lambda\;.
\end{align}
The matrix elements $\lambda_{ij}$ are just the LECs organized according
to their order along the anti diagonals of $\lambda$. For example at N$^3$LO
\be
  \lambda = 
       \begin{pmatrix} 
         C_0 & C_2   & C_4    & C_6 \\
         C_2 & C_{22} & C_{24} &  \\
         C_4 & C_{42} &       &  \\
         C_6 &       &       &        
       \end{pmatrix}
       \;.
\ee 

The loop integrals ${\cal I}_{ij}$ defined in \eqref{def_I}  
depends only on the sum of their indices, i.e. ${\cal I}_{ij}={\cal I}_{2(i+j)}$.
Using the recursive relation \cite{Phillips1998}
\be  \label{I_of_E_recursive}
   {\cal I}_{2k}(E)=2\mu E {\cal I}_{2(k-1)}(E)
     +I_{2k+1}\;,
\ee
with
\be\label{Ik_definition}
    I_{2k+1}=-2\mu \int\frac{d^{3}q}{(2\pi)^{3}}
              F^2(q) q^{2k-2}\;,
\ee
the integral ${\cal I}_{2k}$ can be reduced into the sum 
\be
  {\cal I}_{2k}(E)=(2\mu E)^k {\cal I}_0(E)+\sum_{j=1}^k(2\mu E)^{j-1}I_{2(k-j)+3}
\ee
where
\be\label{I_of_E_definition}
   {\cal I}_0(E)=\int\frac{d^{3}q}{(2\pi)^{3}}
             \frac{F^2(q)}{E+i\varepsilon-\frac{q^2}{2\mu}}\;.
\ee

Explicit expressions for the integrals \eqref{Ik_definition} and \eqref{I_of_E_definition} 
are given for a sharp cutoff in \cite{Epelbaum2018}.
The corresponding expressions for a Gaussian cutoff are given by
\begin{align}\label{I_of_E_gauss}
  {\cal I}_0(E)=-\frac{\mu\Lambda}{(2\pi)^{\frac{3}{2}}}
        +\frac{\mu\sqrt{2\mu E}}{\pi^{\frac{3}{2}}}
        &
           D\left(\sqrt{\frac{4\mu E}{\Lambda^2}}\right)
           \\\notag &
           -i\frac{\mu\sqrt{2\mu E}}{2\pi}e^{-\frac{4\mu E}{\Lambda^2}}
           \;,
\end{align}
and
\be\label{Ik_gauss}
I_{2k+1} = -\frac{\Gamma\left(k+\frac{1}{2}\right) \mu \Lambda^{2k+1}}
                {\sqrt{2}\cdot2^{k+1} \pi^2}\;.
\ee
$D(x)$ is the Dawson function,
\begin{align}
    D(x)&\equiv e^{-x^2}\int_0^xe^{y^2}dy =
    -i \frac{\sqrt{\pi}}{2}e^{-x^2}\text{erf}({i x})\;.
\end{align}

Using these results for Gaussian regulator, the NLO $T$-matrix becomes
\begin{align}\label{T_nlo}
  \frac{1}{T}=e^{\frac{4\mu E}{\Lambda^2}}
        \left(
        \frac{(C_2 I_3-1)^2}
             {C_0+C^2_2 I_5 +\frac{2\mu E}{I_3}\left(1-(C_2 I_3-1)^2\right)}-{\cal I}_0(E)
        \right)\;.
\end{align}

The renormalization conditions for the LECs can be now deduced
by comparing the EFT $T$-matrix \eqref{T_nlo} with the experimental parameters
of the effective range expansion \eqref{eq:t_expansion}.
More specifically, the LO parameter $C_0$ is obtained by inverting the
scattering length equation $a_s=\frac{\mu}{2\pi}T(0)$, yielding the relation
$C_0=C_0(a_s,C_2)$.  
Similarly, the effective range is deduced from the real part of the energy derivative of
the $T$-matrix at zero,
\be\label{reff_eq}
  \reff=\Re\left[ \frac{d}{dE}\left(-\frac{2\pi}{\mu^2 T(E)}\right)\bigg|_{E=0}\;\right].
\ee
Substituting $C_0(a_s,C_2)$ in \eqref{reff_eq} one obtains the following
relation between $\reff$ and $C_2$
\begin{align}\label{reff_nlo}
  \reff &=
  -\frac{4}{\Lambda^2 a_s}
   +\frac{2\pi}{\mu^2}\Re\left[{\cal I}_0'(0)\right]
   \cr & \hspace{-1em}
  -\frac{4\pi}{I_3\mu}\left(\frac{\mu}{2\pi a_s}
   +\Re\left[{\cal I}_0(0)\right]\right)^2
   \left[1-\frac{1}{(C_2I_3-1)^2}\right]\;.
\end{align}
Inspecting the high cutoff limit $\Lambda \to \infty$ of \eqref{reff_nlo}, the 
EFT Wigner bound on $\reff$, for a given scattering length $a_s$,
\be\label{reff_wb}
   \reff \;{\leq}\; \frac{W}{\Lambda}
\ee
is obtained. The Wigner bound parameter $W$ is a positive dimensionless
constant, that may depend on the regulator, and in general also on the order
of our EFT. In the limit $\Lambda \to \infty$, 
Eq. \eqref{reff_wb} leads to the unnatural result
$\reff \leq 0$ \cite{Phillips1998}. 

For an EFT at NLO with Gaussian regulator, the explicit expression
$W=8\sqrt{\frac{2}{\pi}}\left[1-\frac{1}{(C_2 I_3-1)^2}\right]$
can be obtained using Eqs. \eqref{Ik_gauss}--\eqref{T_nlo}. 
It is clear that in this case the maximum value of $\reff$ is acheived when
taking $C_2I_3\to\infty$, leading to $W=8\sqrt{\frac{2}{\pi}}$.

Given an experimental value of $\reff$, 
% 
%SAAR: eq \eqref{reff_wb} is relevant only for high cutoffs. if r_eff is finite, and so is the order.
% the cutoff is not large enough to assume this equation. check it with our results
% 
{Eq. \eqref{reff_wb} can be inverted to
yield an upper bound on the cutoff $\Lambda \leq \Lambda_\text{max} = W/\reff$.}
In the following we
would like to study the dependence of the Wigner bound parameter $W^{(n)}$
on the order $n$ of the EFT, and see if $W^{(n)}$ diverges in the limit
$n \to \infty$, removing the Wigner bound for a complete theory, and restoring
renormalization group invariance.
To this end, we shall concentrate on the several first EFT orders, and try to
infer the general behavior. 

Using Eqs. \eqref{eq:t_expansion} and \eqref{eq:t_matrix} 
the scattering length $a_s$ and the effective range $\reff$ can be expressed through the
relations
\be\label{eq:a_scat}
  a_s=\frac{\mu}{2\pi}\tau_{00}(0)
\ee
and
\be\label{eq:r_eff}
  \reff=\frac{\tau'_{00}(0)+2\mu
    \left(\tau_{01}(0)+\tau_{10}(0)\right)-\frac{8\pi a_s}{\Lambda^2}}{2\pi a_s^2}\;.
\ee
The $\tau$ matrix, Eq. \eqref{tau_solution}, depends on all LECs 
of order $n$ or smaller. After inverting Eq. \eqref{eq:a_scat}, $\reff$ in \eqref{eq:r_eff}
is a function of the scattering length $a_s$ and all the LECs but $C_0$.
Naively, in order to extract $W^{(n)}$ one should take the limit of
Eq.~\eqref{eq:r_eff} at high cutoffs and then search for the maximum of
$\reff$ over the LECs parameter space $\{C_2,C_4,\ldots\}$.
We managed to follow this procedure analytically up to order $n=2$,
i.e. N$^2$LO. Beyond that point this approach becomes impractical, due to
the inversion operation in Eq. \eqref{tau_solution}. Instead, noting that
at NLO, Eq. \eqref{reff_nlo}, the maximum was obtained in the limit
$C_2I_3\to\infty$, we introduced the dimensionless LECs
\begin{align}
    C_{pq}=\frac{\tilde C_{pq}}{\mu\Lambda^{p+q+1}}\;,
\end{align}
Now, following the NLO example and searching for a maximum $\reff$, we impose
the condition $C_{pq} I_{p+q+1}\to\infty$ by setting $\tilde C_{pq} = \Lambda$.
With this condition, it is relatively easy to obtain an analytic expression
for the Wigner bound parameter
$W^{(n)}=\lim\limits_{\Lambda\to\infty} \Lambda\, \reff$.
Of course, this prescription does not guarantee the maximization of $\reff$, 
but it sets a lower bound. Nevertheless, searching numerically for
the maximum of $\reff$  up to N$^4$LO we obtained exactly the same results.
We applied this procedure up to EFT order $n=9$ and $n=10$, for a Gaussian and a
sharp regulator correspondingly. The first six $W^{(n)}$ results for both regulators are presented in
Table~\ref{tab:W_values}.
\begin{table}[ht!]
  \caption{The Wigner bound parameter $W^{(n)}$ as a function of the EFT order   
  $n$ for $n=1,2,\ldots 6$. The expressions are presented for both
    a Gaussian and a sharp regulators.}
  \label{tab:W_values}
  \centering
  \begin{ruledtabular} 
    \begin{tabular}{ccccccc}
      Order &  1 & 2 & 3 & 4 & 5 & 6 \\
      \hline
      {Gaussian} &  $8\sqrt{\frac{2}{\pi }}$    & $\frac{32 \sqrt{\frac{2}{\pi }}}{3}$ 
      & $\frac{64 \sqrt{\frac{2}{\pi }}}{5}$    & $\frac{512 \sqrt{\frac{2}{\pi }}}{35}$ 
      & $\frac{1024 \sqrt{\frac{2}{\pi }}}{63}$ & $\frac{4096 \sqrt{\frac{2}{\pi }}}{231}$
      \\
      {Sharp} & $\frac{16}{\pi }$ & $\frac{256}{9 \pi }$ & $\frac{1024}{25 \pi }$ 
      & $\frac{65536}{1225 \pi }$ & $\frac{262144}{3969 \pi }$ & $\frac{4194304}{53361 \pi }$
  \end{tabular}
  \end{ruledtabular} 
\end{table}

We note that for the Gaussian regulator the results in Table~\ref{tab:W_values} (and those of higher order)
follow the pattern
\be\label{Wn_gauss}
  W^{(n)}_{\text{Gauss}}=4\sqrt{2}\frac{\Gamma\left(n+1\right)}{\Gamma\left(n+\frac{1}{2}\right)}\;
\ee
whereas the results for the sharp cutoff follow the pattern
\be\label{Wn_sharp}
  W^{(n)}_{\text{Sharp}}=4\left(
    \frac{\Gamma\left(n+1\right)}{\Gamma\left(n+\frac{1}{2}\right)}
  \right)^2\;.
\ee
Making the conjecture that these relations hold for any EFT order $n$, and
utilizing Stirling's formula, we hypothesize that at high EFT orders 
\be\label{Wn_gauss_limit}
  \lim\limits_{n\to\infty}W^{(n)}_{\text{Gauss}}=4\sqrt{2n}\;,
\ee
and
\be
  \lim\limits_{n\to\infty}W^{(n)}_{\text{Sharp}}=4n \;.
\ee

If holds, this conjecture implies that $W^{(n)}$ diverge in the limit
$n \to \infty$ defeating the Wigner bound for a complete theory, and restoring
renormalization group invariance for an EFT with positive effective range.

\begin{figure}
  \centering
  \includegraphics[width=0.5\textwidth,page=1]{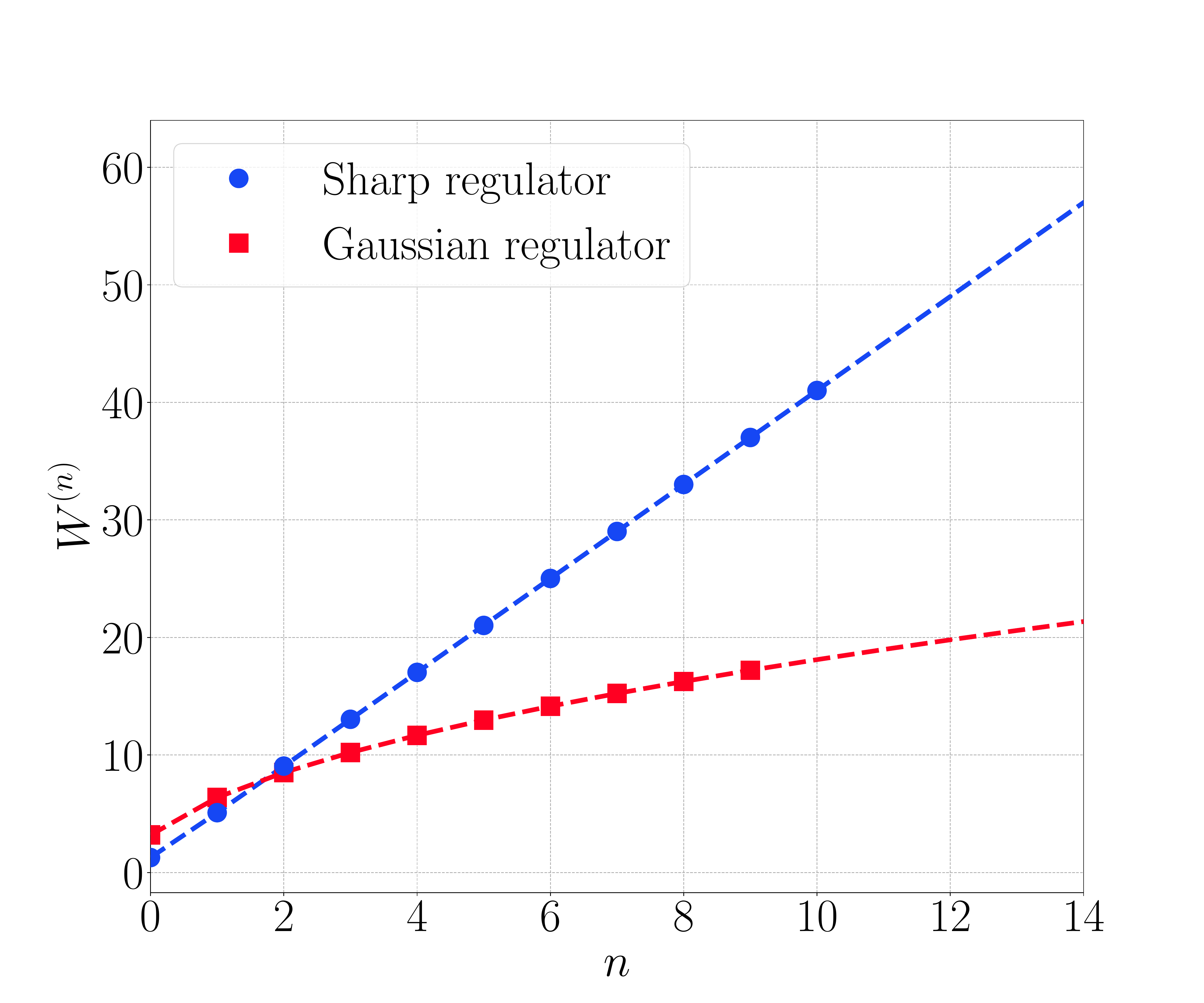}
  \caption{The Wigner bound parameter $W^{(n)}$ as a function of the EFT order   
    $n$. Red squares are results for a Gaussian regulator, blue circles denote 
    a sharp regulator.
    %Dots are the values presented in Table~\ref{tab:W_values},
    The curves are our conjectures Eqs. \eqref{Wn_gauss} and \eqref{Wn_sharp}.}
  \label{fig:max_r_eff_high}
\end{figure}

The difference between the two regulators, displayed in
Fig.~\ref{fig:max_r_eff_high}, is quite striking. It can be viewed as another
indication that short distance physics 
coming from loops is important in non-perturbative EFT. 
Consequently, regulators with different high momentum 
behavior can yield different results.

% #######################
\section{Fixing the LECs}
% #######################

For the EFT to reproduce the available experimental data, the LECs should be 
fitted to some selected observables. For contact EFT it is natural to fit the
LECs to the effective range expansion parameters, Eq. \eqref{eq:t_expansion}. 
In the previous section we have discussed the limited case of fitting the
LECs to reproduce $a_s,\reff$. 
In this section we elaborate more on the fine details of the fitting procedure,
aiming to demonstrate that our Wigner bound parameters $W^{(n)}$ hold also for
the general case where we fit the LECs to reproduce also the shape parameters.

In principle, choosing the best LECs demands nothing but a simple $\chi^2$
minimization. However, iterating the potential to all orders, the scattering
$T$-matrix becomes a non linear function of the LECs. 
As a result, the parameter space $\{C_{pq}\}$ contains multiple minima, that
are equivalent in $\chi^2$. The number of these minima might grow with the 
EFT order.

\begin{figure}
    \centering
    \includegraphics[width=0.5\textwidth,page=3]{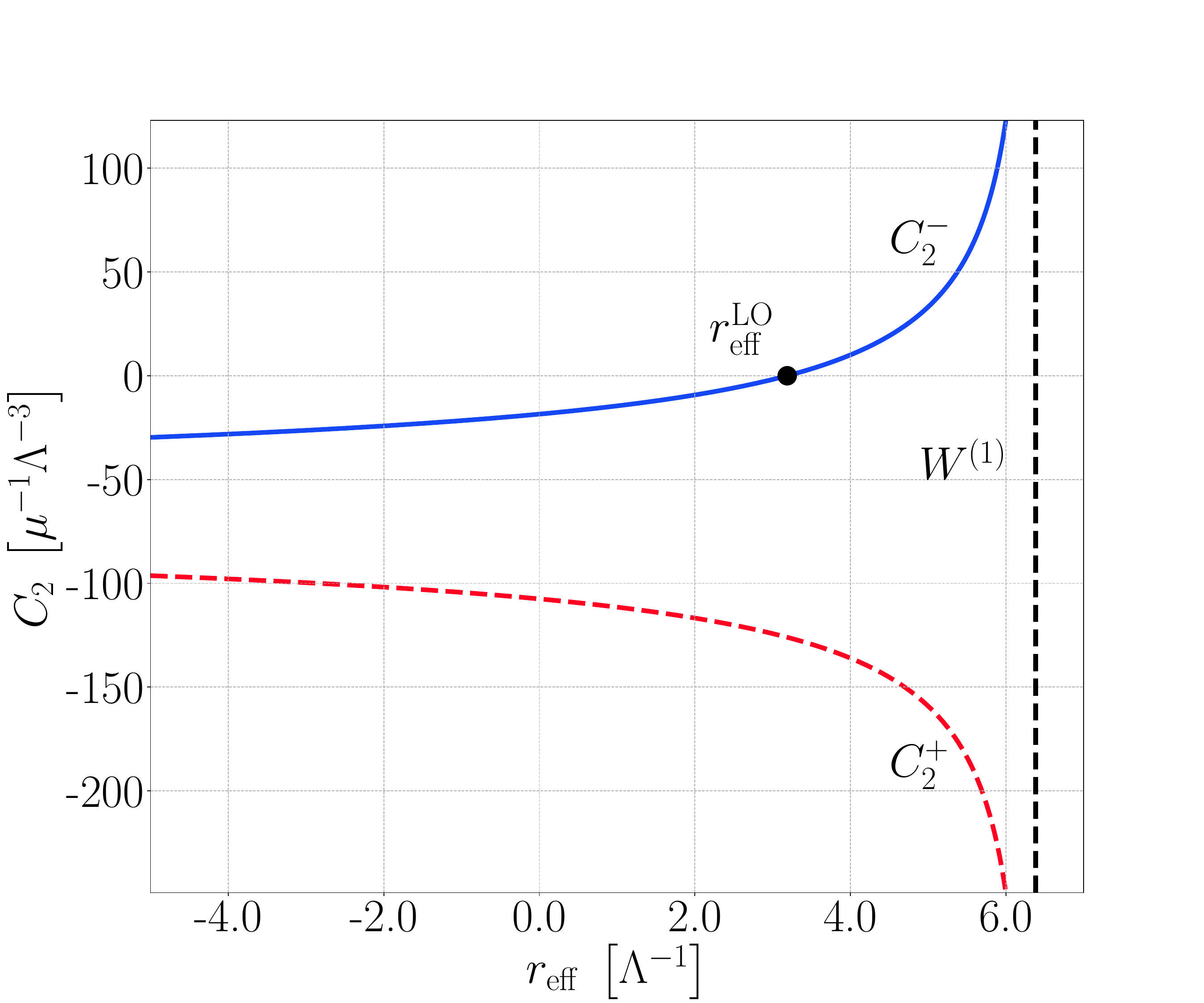}
    \caption{$C_2^{\pm}$ in NLO EFT versus the effective range. 
      Upper branch (blue line) - $C_2^-$, lower branch (red, dashed) - $C_2^+$.
      Vertical dashed line - the Wigner bound at NLO. 
      Black dot - $\reff$ obtained at LO for $a_s\Lambda=10^3$.}
    \label{fig:c2_2_sols}
\end{figure}

To understand the problem, let us reconsider the bosonic EFT of
Eq.~\eqref{eq:lagrangian} at NLO, and choose again $a_s$ and $\reff$ as the
fitting observables. 
Fixing the cutoff $\Lambda$ and solving the Lippmann-Schwinger 
equation, one obtains a closed form expression for $\reff=\reff(C_2)$,
Eq.~\eqref{reff_nlo}, and thus just need to invert it to get $C_2=C_2(\reff)$,
\begin{align}
  C_2^{\pm}&=\frac{1}{I_3}\left[
    1 \pm \left(
      1-\frac{\Re[{\cal I}_0'(0)]-\frac{\mu^2\reff}{2\pi}
       -\frac{2\mu^2}{\Lambda^2\pi a_s}}{\frac{2\mu}{I_3}
       \left(\frac{\mu}{2\pi a_s}+\Re[{\cal I}_0(0)]\right)^2}
    \right)^{-\frac{1}{2}}
    \right]\;.
\end{align}
Studying this expression we first of all note that for the LECs to be real,
ensuring a real action, the expression in the root must be positive. This
condition is nothing else but the Wigner bound. It can also be seen that for
a positive root there exists two solutions for $C_2$.
Using the fact that $I_{2k+1}<0$, Eq. \eqref{Ik_gauss}, we conclude that 
\begin{align}
  C_2^-&\in(\frac{1}{I_3},\infty)\\
  C_2^+&\in(-\infty,\frac{1}{I_3})<0\;.
\end{align}
Hence, only the $C_2^-$ branch contains the zero, i.e. only the minus solution
can be thought of as a continuity of the LO theory. Stated differently,
suppose that the LO completely describes the underlying theory (such as the
case for a delta potential), i.e. it reproduces all the low energy observables,
and specifically the effective range. 
In this situation we expect the NLO theory to be equivalent to the LO theory, 
i.e. $C_2=0$. This solution is not accessible by the $C_2^+$ branch, we
therefore conclude that $C_2^-$ is the physical branch. 
This situation is presented graphically in Fig.~\ref{fig:c2_2_sols}.
% 
% SAAR: It did not happen in N2LO. Of course, I believe that it will happen again in higher orders
% 
The above situation can repeat itself at higher EFT orders. As analytical
calculations become much harder with increasing $n$, one must resort to
numerical computations. In order to identify numerically the physical
solution we suggest the following strategy:
For a given cutoff $\Lambda$, start at LO and fix $C_0$ to reproduce one low
energy observable, say $a_s$. Proceed to NLO and start the search with
$C_0=C_0^{\text{LO}}$ and $C_2=0$. Now change $C_2$ slowly until the theoretical
effective range $\reff^{\text{NLO}}$ matches the experimental value $\reff$.
Make sure that in the process the values of $C_0,C_2$ do not jump from one
solution branch to another. Repeat the process with each EFT order. 
This process can be visualized in Fig. \ref{fig:c2_2_sols}, as follows:
we start from the LO solution $C_2=0$, black dot,
and move slowly along the $C_2^-$ branch until we reproduce the experimental
$\reff$.

% #########################
\section{Numerical example}
\label{numerical_example}
% #########################

\begin{figure}
    \centering
    \includegraphics[width=0.5\textwidth,page=2]{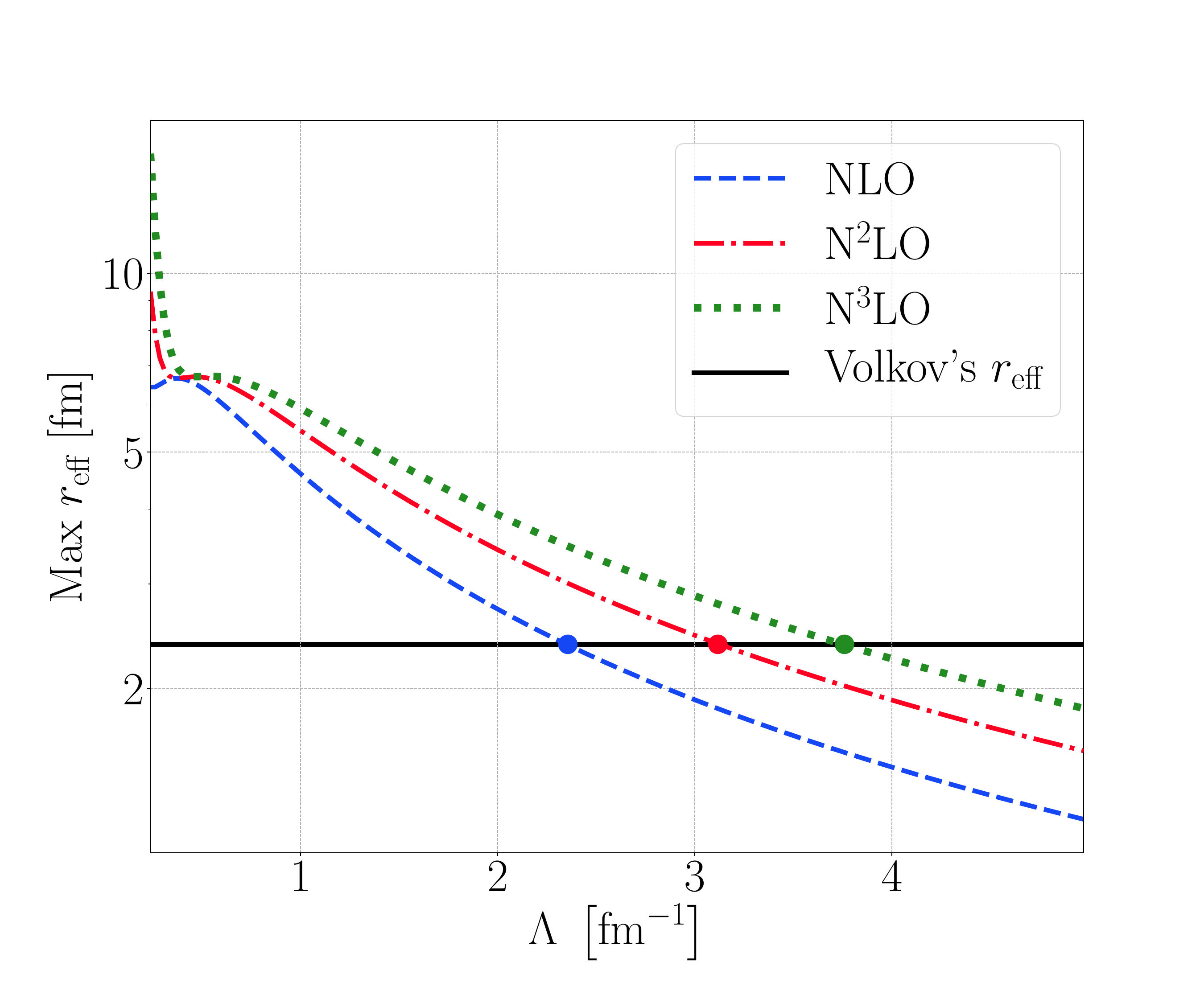}
    \caption{The maximal value of $\reff$ as a function of the cutoff for
      different EFT orders. $a_s$ was fixed to fit the Volkov potential.
      The calculations were done with a Gaussian regulator. The dots shows
      the actual $\reff$ value for the Volkov potential.}
    \label{fig:max_r_eff_low}
\end{figure}

In section \ref{Effective range in various EFT orders} we aimed 
at getting the Wigner bound 
as a function of the EFT order $n$
while ignoring the matching between the 
LECs and all physical observables but $a_s,\reff$. 
Our conclusion was that $W^{(n)}$ increases indefinitely with EFT order. 
In this section we want to verify this observation through a concrete numerical example
where we fit not only $a_s,\reff$ but also the leading shape parameter. 
To this end we consider a synthetic example where
our underlying theory consists of
bosonic nucleons, i.e. bosons with the nucleon mass, 
that interact via the 2-body Volkov potential.
To simplify the numerical work, we limit ourselves in this section
to a Gaussian regulator.

The Volkov potential is given by
\be
  V(r)=V_Re^{-\frac{r^2}{R_1^2}}+V_Ae^{-\frac{r^2}{R_2^2}}, 
\ee
where $V_R=144.86\;\text{MeV}, \;V_A=-83.34\; \text{MeV}$, 
$R_1=0.82\;\text{fm}$, and $R_2=1.6 \;\text{fm}$.
Its leading effective range parameters are
$a_s=10.08\,\text{fm}$, $\reff=2.37\,\text{fm}$ and $S_2=0.43\,\text{fm}^3$. 
The 2-body binding energy is $E_2=-0.545\,\text{MeV}$.

We start by calculating the Wigner bound, i.e. the maximal effective range
as a function of the cutoff $\Lambda$
assuming that the
Volkov potential is our underlying theory, and fixing $a_s$ to $10.08\,\text{fm}$.
The results are presented in Fig. \ref{fig:max_r_eff_low}.
Defining $\Lambda^{(n)}_\text{max}\equiv W^{(n)}/\reff$ as the highest cutoff that can be taken 
at order $n$ while still reproducing the experimental effective range,
we see that for the Volkov case $\Lambda^{(n)}_\text{max}=2.3,\,3.1,\,3.8\;\text{fm}^{-1}$
for $n=1,2,3$. Comparing these numbers with our analytical prediction
Eq. \eqref{Wn_gauss_limit} we see that indeed as expected
$\Lambda^{(n+1)}_\text{max}/\Lambda^{(n)}_\text{max}\approx \sqrt{(n+1)/n}$.

Now we limit our attention to $n=2$, i.e. EFT at N$^2$LO, and utilize the
effective range expansion parameters to order $p^4$ to fit the LECs. 
We note that at this order there are 3 observables $a_s,\reff,S_2$, but
4 LECs. Considering only the on-shell 2-body $T$-matrix, it is well established
that there is a one-to-one correspondence
between the effective range expansion and contact EFT 
while some of the LECs become redundant \cite{Beane2001,vanKolck1999}. 
These LECs encode information on the off-shell physics and therefore cannot be 
renormalized via scattering data.
It follows that if we focus on the 2-body sector at N$^2$LO one of the LECs will
remains free \cite{Arzt1995}. In the following we will utilize this parameter
as a measure for how much freedom remains 
after renormalization at a specific cutoff value. 

\begin{figure}
    \centering
    \includegraphics[width=0.5\textwidth,page=5]{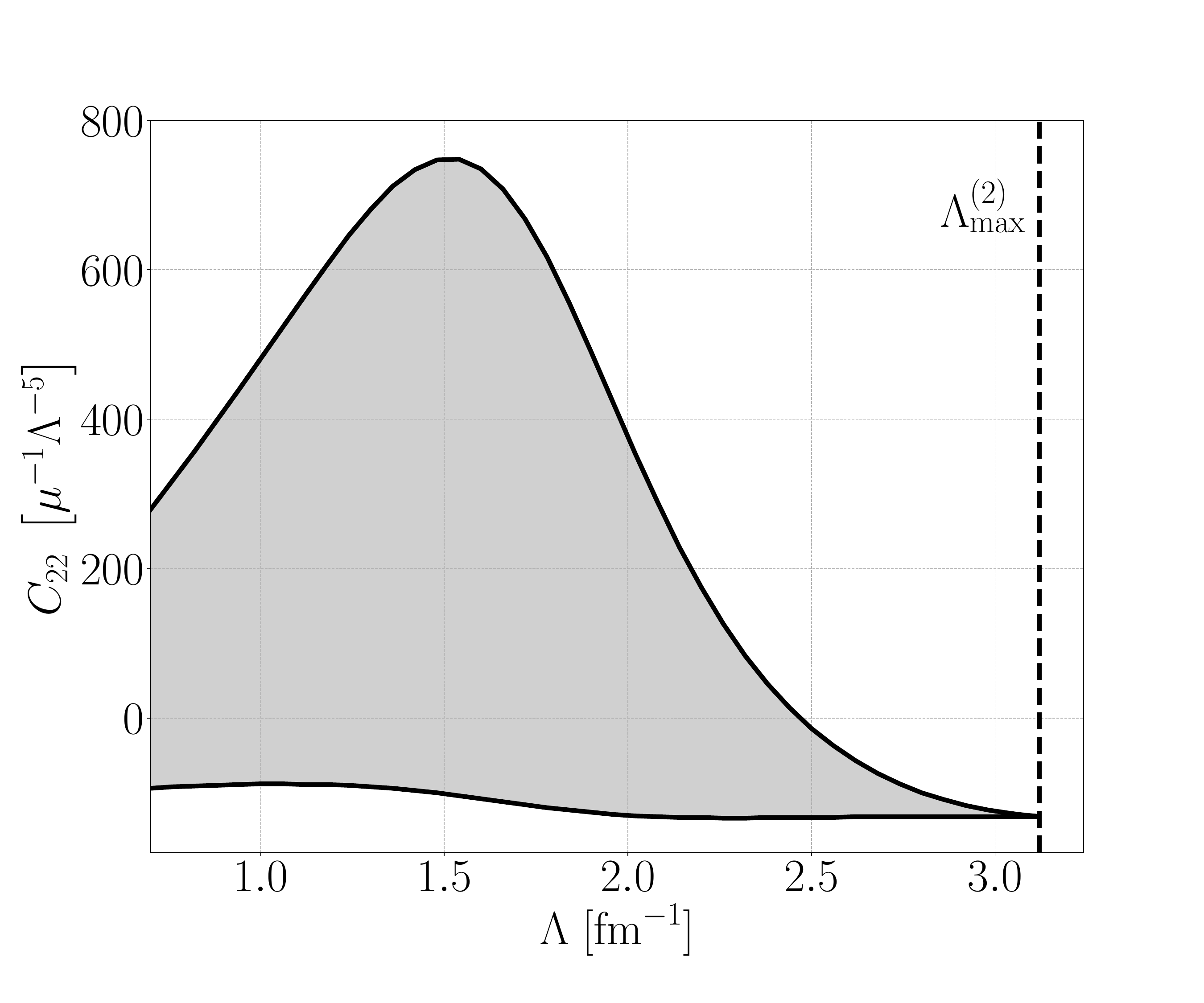}
    \caption{EFT at N$^2$LO. The permissible values of the LEC $C_{22}$ as a 
      function of the cutoff - shaded area. 
      Vertical dashed line - the highest cutoff reproducing $\reff$ at N$^2$LO. 
      The LECs $C_0,C_2,C_4$ are fitted to reporduce the Volkov potential $a_s,\reff,S_2$.}
    \label{fig:c4_interval}
\end{figure}

In practice we have derived an analytic expression for the LECs $C_0,C_2,C_4$
that depends on $a_s,\reff, S_2$ and $C_{22}$, which we kept as a free parameter.
We have found that the permissible values of $C_{22}$ were limited
by the hermiticity condition that $C_0,C_2,C_4$ are real. 
In Fig. \ref{fig:c4_interval} we present this permissible range of $C_{22}$
as a function of the cutoff. 
It can be seen, that close to the critical point 
$\Lambda\longrightarrow\Lambda^{(2)}_\text{max}$ 
the permissible range shrinks to a point, and that it completely 
disappears when $\Lambda > \Lambda^{(2)}_\text{max}$.

\begin{figure}
    \centering
    \includegraphics[width=0.5\textwidth,page=4]{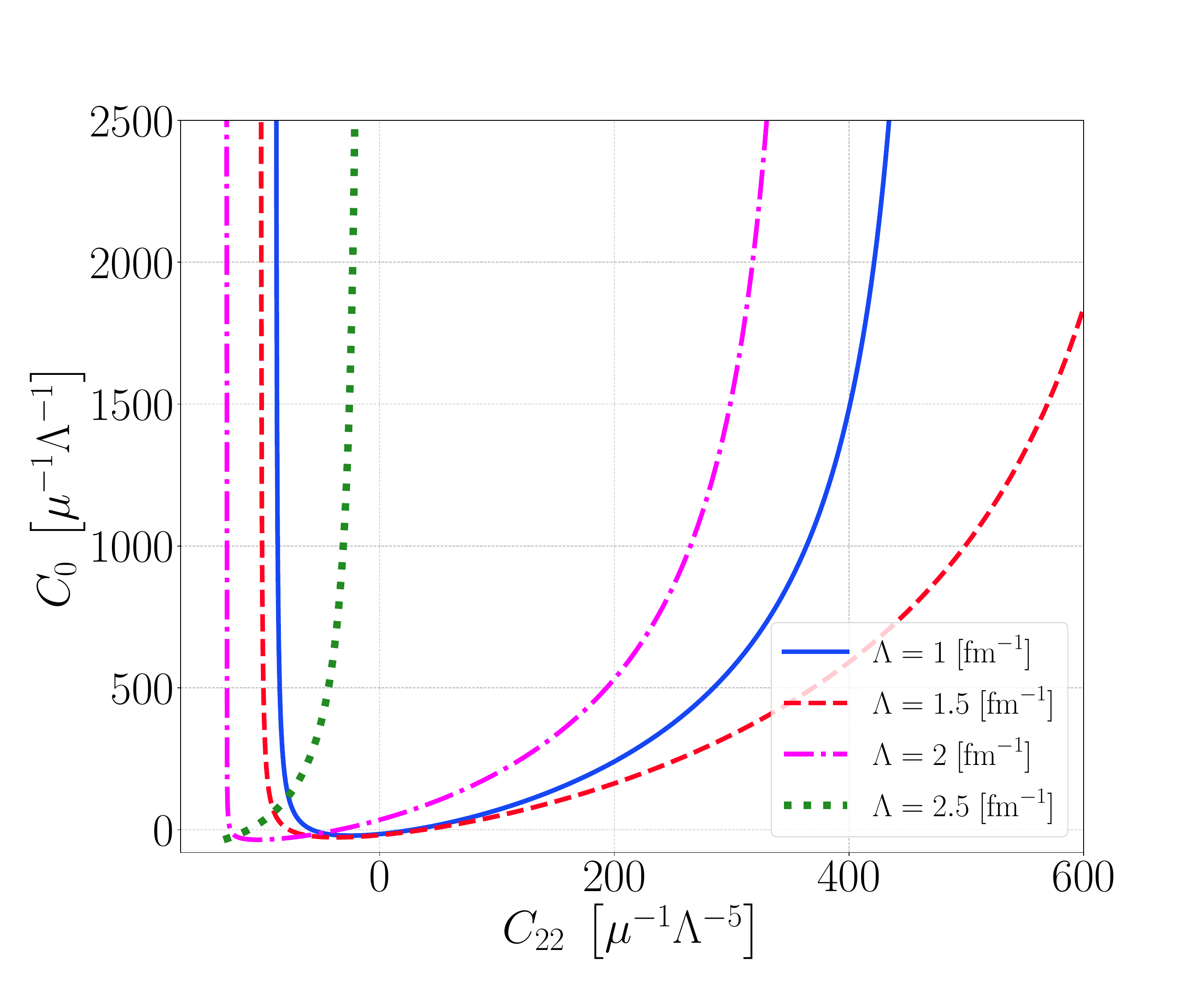}
    \caption{The LEC $C_0$ as a function of ${c}_{22}$ at differect cutoffs.
      The scattering parameters
      $a_s, \reff, S_2$ are constrained by the Volkov potential. 
      Blue (full) line - $\Lambda = 1.0\;\text{fm}^{-1}$,
      red (dashed) line - $\Lambda = 1.5\;\text{fm}^{-1}$,
      purple (dashed-dot) line - $\Lambda = 2.0\;\text{fm}^{-1}$, and
      green (dotted) line - $\Lambda = 2.5\;\text{fm}^{-1}$.}
    \label{fig:c0_vs_t4}
\end{figure}

Trying the behavior of the theory near the edge of the permissible $C_{22}$
range, we explored the relations between $C_{22}$ and the other LECs.
It appears that at the edge of the 
$C_{22}$ region the LECs possess a singular point, i.e. the
renormalization conditions for $C_0(C_{22}),C_2(C_{22}),C_4(C_{22})$ diverge. 
It follows that as the freedom in $C_{22}$ decreases, the relations between the LECs
become more radical. To illustrate this point we plot in Fig. \ref{fig:c0_vs_t4}
the LEC $C_0$ as a function of $C_{22}$ for different cutoffs. 

\begin{figure}
    \centering
    \includegraphics[width=0.5\textwidth,page=6]{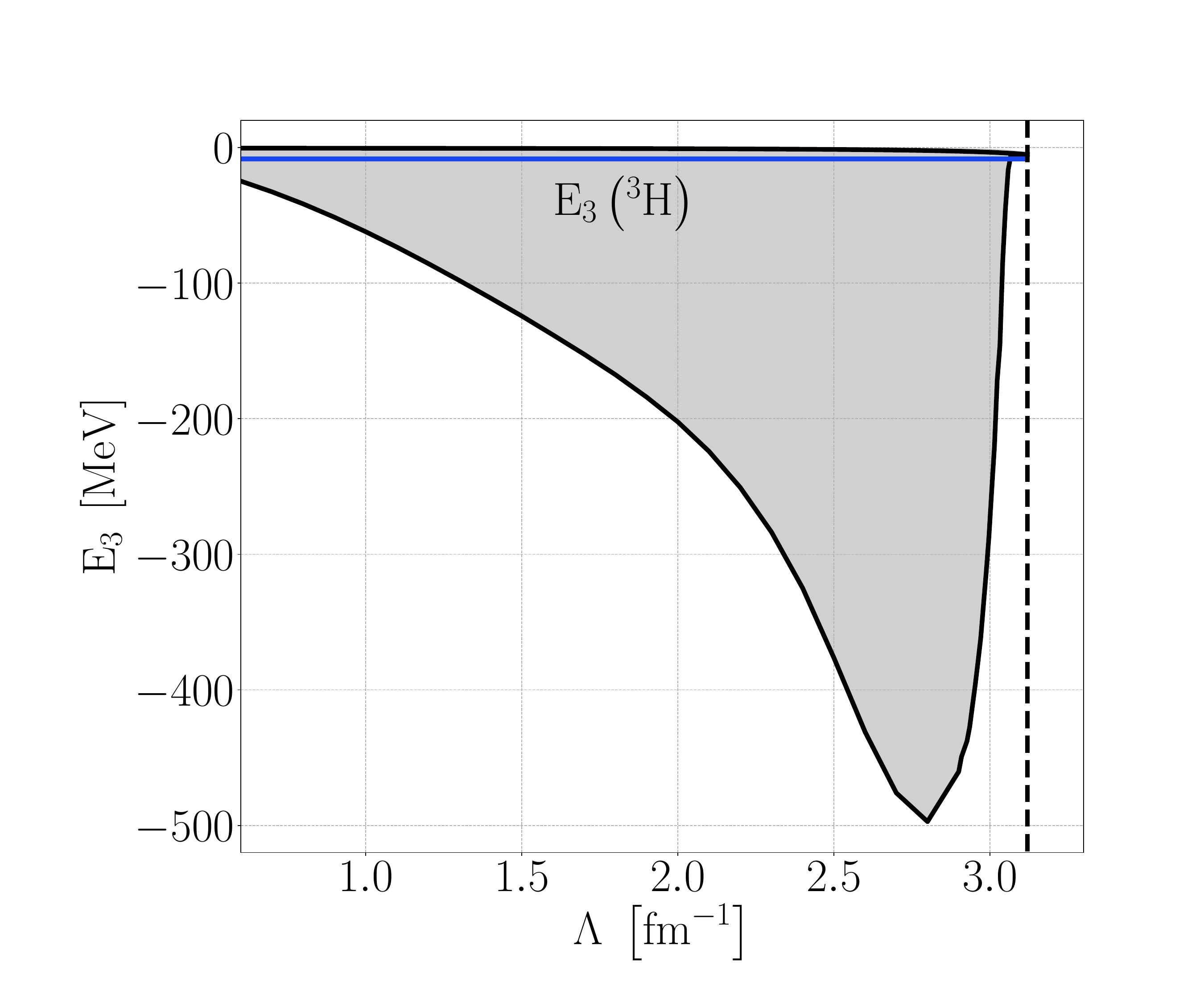}
    \caption{The triton's binding energy $E_3$ as a function of the cutoff. 
      Blue line - the exact value for the Volkov potential. Shaded 
      gray area - range of possible EFT (at N$^2$LO) predictions, reflecting 
      the permissible values of $C_{22}$, see Fig. \ref{fig:c4_interval}.}
    \label{fig:tritium}
\end{figure}

The shrinking of the  permissible $C_{22}$ interval in the limit 
$\Lambda\longrightarrow\Lambda^{(2)}_\text{max}$ 
may lead us to think that in this limit $C_{22}$ becomes redundant.
That is, the range of possible predictions implied from the freedom 
in $C_{22}$ should shrink as well.
To check this hypothesis we have utilized our EFT to calculate the 
triton's binding energy $E_3(\text{Volkov})$, that for the Volkov potential is 
equal to $-8.431\,\text{MeV}$. 
The results of these calculations are presented 
in Fig. \ref{fig:tritium} as a function of $\Lambda$. 
Surprisingly, we have found that up to rather high values of $\Lambda$ the $E_3(C_{22})$
range increases while the permissible interval of $C_{22}$ decreases. 
At low cutoffs, the upper bound on $E_3$ coincides with the 2-body threshold,
and it decreases slowly as $\Lambda\longrightarrow\Lambda^{(2)}_\text{max}$.
In contrast, the lower bound drops dramatically, following the standard path of 
the Thomas collapse \cite{Thomas1935a} up to $\Lambda \approx 2.7\,\text{fm}^{-1}$. 
Above this point the collapse stops, and near the critical point
$\Lambda\approx \Lambda^{(2)}_\text{max}$ the values of $E_3$ become comparable
to the exact binding energy $E_3(\text{Volkov})$.

\begin{figure}
    \centering
    \includegraphics[width=0.5\textwidth,page=7]{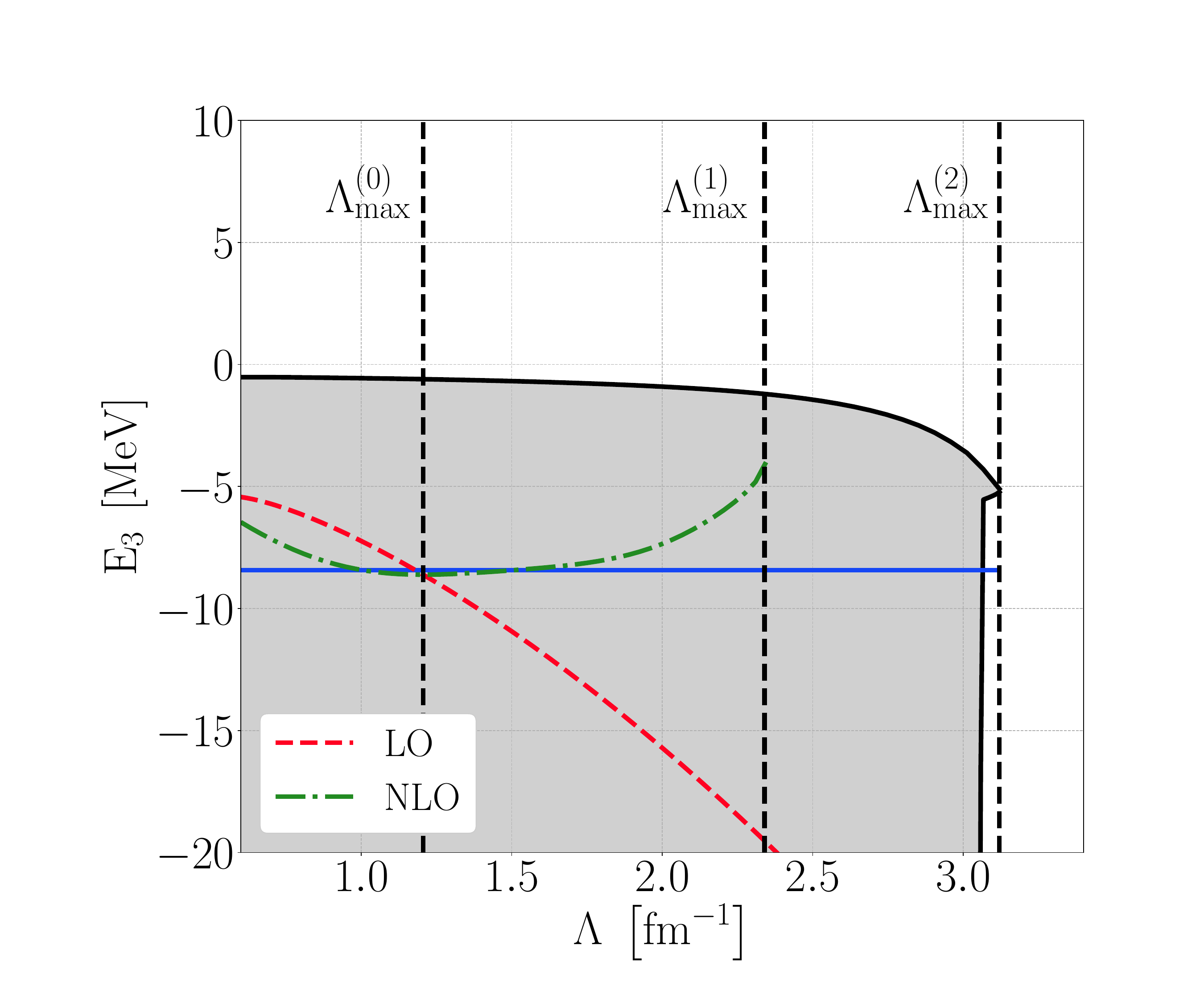}
    \caption{The triton's binding energy $E_3$ as a function of the cutoff, 
      like in Fig. \ref{fig:tritium}, with the addition of the LO, and NLO
      results.
    }
    \label{fig:triton_order_by_order}
\end{figure}

To better understand this result and its consequences, we have calculated
the triton's binding energy also at LO, and NLO, see
Fig. \ref{fig:triton_order_by_order}.
From the figure it can be seen that if the LO calculations are performed 
at $\Lambda=\Lambda^{(0)}_{\max}$,
reproducing $\reff$, the resulting $E_3(\text{LO})$ is rather close to $E_3(\text{Volkov})$.
For NLO this point is extended into a range
over which $E_3(\text{NLO})\approx E_3(\text{Volkov})$, 
but as $\Lambda\to\Lambda^{(1)}_{\max}$ the energy deviates and at the critical point
$E_3\approx -5~\text{MeV}$. We note that at N$^2$LO we can renormalization $C_{22}$ to 
reproduce $E_3(\text{Volkov})$, thus eliminating the 3-body force. 
This we can do up to about
$\Lambda = 3~\text{fm}^{-1}$, the point where the exact result is excluded from the 
permissible $E_3$ region. From this point on $E_3$ will move along the lower bound 
of the region
until at $\Lambda^{(2)}_{\max}$ it will reach a value $E_3\approx -5~\text{MeV}$. 
Following a pattern similar to NLO.

%\newpage
%=============================================================================
\section{Summary}
%=============================================================================
For a non-perturbative contact EFT we analyzed 
the evolution of the Wigner bound parameter $W^{(n)}$
with the EFT order $n$.
Considering sharp and Gaussian regulators, 
we have found, up to $n=10$ and $n=9$ respectively, an analytic lower
limit for $W^{(n)}$. 
This limit is regulator specific. 
From this analysis, we have concluded that the Wigner bound loosens
with increasing EFT order, and conjectured for these regulators
the general dependence of $W^{(n)}$ on $n$.
If holds, this conjecture implies that $W^{(n)}$ diverge in the limit
$n \to \infty$ defeating the Wigner bound, and restoring
renormalization group invariance for a complete theory.

Verifying our results with a concrete numerical example we have demonstrated at
N$^2$LO that our conclusions hold after full renormalization procedure, 
as long as at least one LEC is utilized to maximize the cutoff. 
Studying the 3-body system with this example, 
we have found that limiting the permissible range 
of cutoffs by the Wigner bound, we avoid the
Thomas collapse, and don't need to promote the 3-body force to LO.
If proven for the general case this observation might be of practical
importance,
as a 3-body force appearing at LO, and 
4-body force appearing at NLO, are a huge liability from a computational 
many-body perspective.

The implications of the current observation on the nuclear \nopieft,
on $p$-wave interaction, and on $\chi$EFT
call for further study.
%Another point that needs further clarification is the extent to which
%our conclusions are regulator shape dependent.

%=============================================================================
\begin{acknowledgments}
%######################
  We would like to thank Bira van Kolck, Chen Ji, Sebastian K{\"o}nig, 
  and Ronen Weiss
  for useful discussions and communications during the preparation of this
  manuscript.
  This work was supported by the Israel Science Foundation (grant number 1308/16).
  SB was also supported by Israel Ministry of Science and Technology (MOST).
\end{acknowledgments}

% #########################

\end{document}